\documentclass[%
reprint,
nolongbibliography,
 amsmath,amssymb,
 aip,
floatfix,
]{revtex4-2}

\usepackage{comment}

\usepackage{lgreek}
\usepackage{bookmark}
\usepackage{acronym}
\usepackage[nameinlink,english]{cleveref}

\hypersetup{
	pdfstartview=FitV, 
}

\usepackage[utf8]{inputenc}
\usepackage[T1]{fontenc}
\usepackage{mathptmx}
\usepackage{etoolbox}

\usepackage{subfig}
\usepackage{wasysym}
\usepackage{stmaryrd}
\usepackage{amsfonts}
\usepackage{xcolor}
\usepackage{multirow}

\usepackage{graphicx}
\usepackage{dcolumn}
\usepackage{bm}
\usepackage{hyperref}

\allowdisplaybreaks

\makeatletter
\def\@email#1#2{%
 \endgroup
 \patchcmd{\titleblock@produce}
  {\frontmatter@RRAPformat}
  {\frontmatter@RRAPformat{\produce@RRAP{*#1\href{mailto:#2}{#2}}}\frontmatter@RRAPformat}
  {}{}
}%
\makeatother

\begin{document}


\title[Extension of SPP framework relating the second Piola-Kirchhoff stress tensor to the Green-Lagrange strain tensor]{Extension of Sequence of Physical Processes framework relating the second Piola-Kirchhoff stress tensor to the Green-Lagrange strain tensor}

\author{Louis-Vincent Bouthier}
\email{louis.bouthier@etu.minesparis.psl.eu}
\author{Romain Castellani}%
\author{Rudy Valette}
\affiliation{%
 Groupe CFL, CEMEF, Mines Paris, PSL Research University, 1 Rue Claude Daunesse, 06904 Sophia Antipolis, France
}%

%




\date{\today}

\begin{abstract}
An extension of the Sequence of Physical Processes using geometrical corrections of the Piola-Kirchhoff stress tensor and the Green-Lagrange strain tensor is addressed. More precisely, the usual Sequence of Physical Processes omits some geometrical non linearities that appear when the deformation becomes large. With this extension, geometrical corrections are added and let the opportunity to study rheological non linearities. Application on two famous classical viscoelastic models, namely the linear Maxwell model and the linear Kelvin-Voigt model, helps to understand how some complex behaviours may be rationalised to better understand the behaviours after some corrections. 

\end{abstract}

\maketitle

\section{Introduction}

General rheological models started to be created when continuum mechanics needed traditional ones to close their equations \cite{Landau1959,LandauLifshitz1987,Barber2004,Bird2002}. In fact, rheology as a discipline started years before the complete formalisation of continuum mechanics and relied mainly on empirical observations\cite{vonMises1913,Tresca1864,Bingham1922,Ostwald1925,Herschel1926a,Maxwell1867,Kelvin1890,Voigt1890}. At the end of the twentieth century, studies began to rationalise the construction of rheological models to comply with thermodynamics for instance\cite{Halphen1975,Bird2002}. These framework were very important to properly account for physical phenomena happening in certain materials. Besides, most of the usual rheological models were first built in small deformations framework to avoid, at first, non linearities. However, recent studies\cite{Wang2022b,Klages2022,Lopez2021} have demonstrated the interest of large deformations study to properly understand the behaviour of some materials. The proper theoretical interpretation must take into account some geometrical effect when materials evolve due to forces, displacement or other solicitations. For instance, in continuum mechanics, the Eulerian framework and the Lagrangian framework helps to distinguish between the current configuration and the initial configuration in terms of strains or in terms of stresses\cite{Bergman2011,Bhatti1987}. Even if this type of consideration may be found in certain studies, it is not the case for  all rheological frameworks. 

In this paper, the Sequence of Physical Processes designed by \cite{Rogers2017,Donley2019,Rogers2011,Rogers2012a,Rogers2012b} is analysed in the perpective of large deformations. The original development is indeed made in a \emph{linear} way without accounting for geometrical corrections at large deformations. The extension of this framework will be addressed through the use of the second Piola-Kirchhoff stress tensor and the Green-Lagrange strain tensor, which are thermodynamics conjugates, and allow the analysis to come back to the initial configuration. 

In \cref{sec:Derivation}, the theoretical development of the extended framework will be presented. Then, in \cref{sec:Applications}, two application cases are studied with the two classical linear viscoelastic models, namely, the Maxwell model\cite{Maxwell1867} and the Kelvin-Voigt model\cite{Kelvin1890,Voigt1890}. Afterwards, some remarks and warning are highlighted in \cref{sec:Experiments}. Finally, some conclusions are drawn in \cref{sec:Conclusion}. 

\section{Derivation of the extension} \label{sec:Derivation}

To recall the usual sequence of Physical Processes presented in \cite{Rogers2017,Donley2019,Rogers2011,Rogers2012a,Rogers2012b}, it is possible via the Fr\'enet-Serret frame to find a transient viscoelastic modulus $G_t$ relating the shear stress $\sigma$ to the shear strain $\varepsilon$ and the shear strain rate $\dot{\varepsilon}$. The whole demonstration relies on the fact that there is a linear relationship in the frequency domain between the stress tensor $\boldsymbol{\sigma}$ and the strain tensor $\boldsymbol{\varepsilon}$ of the form $\boldsymbol{\sigma}=\mathbb{C}\boldsymbol{\varepsilon}$ with $\mathbb{C}$ a fourth order tensor gathering the material properties. Acknowledging the power of such tools, it is interesting to question the relevance of these last assumptions. For instance, when increasing the amplitude of the oscillation of the shear strain, rheological non-linearities may appear, as well as geometrical non-linearities. Then, it may be interesting to decorrelate both previous effects to study only the rheological non-linearities. Such a framework exists in continuum mechanics and is related to hyperelastic materials. Precisely, the Cauchy stress tensor $\boldsymbol{\sigma}$ and the linear strain tensor $\boldsymbol{\varepsilon}$ are properly defined in the deformed configuration of a material. When the latter undergoes large deformation, the deformed configuration becomes pretty different from the reference configuration. Therefore, it is possible to create two quantities that are defined in the reference configuration and allow to connect stresses and deformations: namely, the second Piola-Kirchhoff stress tensor $\boldsymbol{S}$ and the Green-Lagrange strain tensor $\boldsymbol{e}$. To recall the construction of such quantities, the transformation from the reference configuration to the deformed configuration is determined by the deformation gradient of the displacement $\boldsymbol{F}$. Using some differential transformation of distances, areas and volumes, the deformations from the reference configuration are properly measured by the Green-Lagrange strain tensor defined by 
\begin{equation}
    \boldsymbol{e}=\frac{1}{2}\left(\boldsymbol{F}^T\boldsymbol{F}-\boldsymbol{I}\right) \label{eq:GreenLagrange}
\end{equation}
with $\cdot^T$ the transpose of a tensor and $\boldsymbol{I}$ the identity tensor. Simultaneously, the forces in the deformed configuration acting on areas in the deformed configuration are given by the Cauchy stress tensor $\boldsymbol{\sigma}$ and can be pulled back to forces in the reference configuration acting on areas in the reference configuration through the second Piola-Kirchhoff stress tensor $\boldsymbol{S}$ defined by 
\begin{equation}
    \boldsymbol{S}=J\boldsymbol{F}^{-1}\boldsymbol{\sigma}\boldsymbol{F}^{-T}\label{eq:SecondPiolaKirchhoff}
\end{equation}
with $J=\det\left(\boldsymbol{F}\right)$. In the case of small deformation, $J\approx1$\footnote{It may be interesting to comment that in the case of incompressible deformations, $J=1$ at every time}, $\boldsymbol{F}\approx\boldsymbol{I}+\boldsymbol{\nabla u}$ with $\boldsymbol{u}$ the displacement, thus the Green-Lagrange strain tensor $\boldsymbol{e}$ is equal to the usual linear strain tensor $\boldsymbol{\varepsilon}$ and the second Piola-Kirchhoff stress tensor $\boldsymbol{S}$ is equal to the usual Cauchy stress tensor $\boldsymbol{\sigma}$. However, in the case of large transformations, the Green-Lagrange strain tensor and the second Piola-Kirchhoff stress tensor account for rotations and large deformations. Relating to the hyperelastic materials framework, it can be demonstrated that the second Piola-Kirchhoff stress tensor and the Green-Lagrange strain tensor are energy conjuguates meaning that a rheological law involving an expression of $\boldsymbol{S}$ as a function of $\boldsymbol{e}$ or its derivatives is objective and properly defined. In addition, it can be demonstrated\cite{Letallec2019} that the mechanical energy per unit volume in the reference configuration due to the material is equal to $\boldsymbol{S}:\dot{\boldsymbol{e}}$ with $:$ the double contracted product and $\dot{\left(\cdot\right)}$ the time derivative. Therefore, in the Fourier domain, it is straightforward to consider, as a first approach, that it exists a complex fourth order tensor $\mathbb{H}$ reading $\boldsymbol{S}=\mathbb{H}\boldsymbol{e}$.

Now with these previous remarks, it is possible to extend the usual Sequence of Physical Processes method. In this framework, an imposed oscillatory shear strain $\varepsilon=\varepsilon_0\sin\left(\omega t\right)$ with $\varepsilon_0$ the shear strain amplitude and $\omega$ the pulsation drives the deformation of the material. Assuming that the shear strain is imposed in the $xy$ plane in cartesian coordinates and using the definition of the deformation gradient $\boldsymbol{F}$ and \cref{eq:GreenLagrange,eq:SecondPiolaKirchhoff}, one gets 
\begin{align}
    \boldsymbol{F}&=\boldsymbol{I}+\varepsilon\boldsymbol{e}_x\otimes\boldsymbol{e}_y\\
    \boldsymbol{S}&=\boldsymbol{\sigma}-\varepsilon\begin{pmatrix}2\sigma_{xy}-\varepsilon\sigma_{yy} & \sigma_{yy} & \sigma_{yz}\\
\sigma_{yy} & 0 & 0\\
\sigma_{yz} & 0 & 0
\end{pmatrix} \label{eq:StressTensorPiola}\\
2\boldsymbol{e}&=\varepsilon\left(\boldsymbol{e}_{x}\otimes\boldsymbol{e}_{y}+\boldsymbol{e}_{y}\otimes\boldsymbol{e}_{x}\right)+\varepsilon^{2}\boldsymbol{e}_{y}\otimes\boldsymbol{e}_{y}.
\end{align}
Hence, while in the usual Sequence of Physical Processes framework, one can write $\sigma=G^{\prime}\varepsilon+G^{\prime\prime}\dot{\varepsilon}/\omega$ with $\dot{\varepsilon}$ the shear strain rate, here one gets 
\begin{equation}
    S=G^\prime\varepsilon+G^{\prime\prime}\frac{\dot{\varepsilon}}{\omega}+H^\prime\varepsilon^2+H^{\prime\prime}\frac{2\varepsilon\dot{\varepsilon}}{\omega}\label{eq:LinearDecomposition}
\end{equation}
with $S$ any components of $\boldsymbol{S}$ and $G^\prime,G^{\prime\prime},H^\prime,H^{\prime\prime}$ the various moduli of the rheological law which are going to be extended in instantaneous values.
To build these instantaneous values, instead of considering a three dimensions space with a position vector $\boldsymbol{x}=\left(\varepsilon,\dot{\varepsilon}/\omega,\sigma\right)$, let us consider a position vector $\boldsymbol{x}=\left(\varepsilon,\dot{\varepsilon}/\omega,\varepsilon^2,2\dot{\varepsilon}\varepsilon/\omega,S\right)$ and let us try to  build a Fr\'enet-Serret apparatus on this five dimensions space. Following the Gram-Schmidt procedure for the vector family $\left(\mathrm{d}_t^j\boldsymbol{x}\right)_{j\in\left\llbracket 1,5\right\rrbracket}$, we define a family $\left(\boldsymbol{e}_i\right)_{i\in\left\llbracket1,5\right\rrbracket}$ of orthonormal vectors constructing the local instantaneous coordinate system following, for all $j\in\left\llbracket2,4\right\rrbracket$,
\begin{align}
    \boldsymbol{e}_{1}&=\frac{\mathrm{d}\boldsymbol{x}}{\mathrm{d} t}\left\Vert \frac{\mathrm{d}\boldsymbol{x}}{\mathrm{d} t}\right\Vert ^{-1}\\
    \overline{\boldsymbol{e}_j}&=\frac{\mathrm{d}^j\boldsymbol{x}}{\mathrm{d} t^j}-\sum_{i=1}^{j-1}\left(\frac{\mathrm{d}^j\boldsymbol{x}}{\mathrm{d} t^j}\cdot\boldsymbol{e}_i\right)\boldsymbol{e}_i\\
    \boldsymbol{e}_j&=\frac{\overline{\boldsymbol{e}_j}}{\left\lVert\overline{\boldsymbol{e}_j}\right\rVert}\\
    \boldsymbol{e}_5&=\boldsymbol{e}_1\times\boldsymbol{e}_2\times\boldsymbol{e}_3\times\boldsymbol{e}_4.
\end{align}
Following the demonstration in \cite{Rogers2017,Donley2019}, one can find at the end 
\begin{equation}
    \begin{pmatrix}G^\prime_{t}\\
G^{\prime\prime}_{t}\\
H^\prime_{t}\\
H^{\prime\prime}_{t}
\end{pmatrix}=-\frac{1}{e_{5}^{S}}\begin{pmatrix}e_{5}^{\varepsilon}\\
e_{5}^{\dot{\varepsilon}/\omega}\\
e_{5}^{\varepsilon^{2}}\\
e_{5}^{2\varepsilon\dot{\varepsilon}/\omega}
\end{pmatrix}
\end{equation}
with $e_5^i$ where $i\in\left\{\varepsilon,\dot{\varepsilon}/\omega,\varepsilon^2,2\dot{\varepsilon}\varepsilon/\omega,S\right\}$ is the component on the axis $i$ in the 5 dimensions space of the vector $\boldsymbol{e}_5$. If someone wants to completely write the expressions of $G^\prime,G^{\prime\prime},H^\prime,H^{\prime\prime}$ as a function of $\left(\varepsilon,\dot{\varepsilon}/\omega,\varepsilon^2,2\dot{\varepsilon}\varepsilon/\omega\right)$, one gets by conservation of the generated vector spaces in the Gram-Schmidt process,  
\begin{equation}
    \overline{\boldsymbol{e}_{5}}=\frac{\mathrm{d}\boldsymbol{x}}{\mathrm{d}t}\times\frac{\mathrm{d}^{2}\boldsymbol{x}}{\mathrm{d}t^{2}}\times\frac{\mathrm{d}^{3}\boldsymbol{x}}{\mathrm{d}t^{3}}\times\frac{\mathrm{d}^{4}\boldsymbol{x}}{\mathrm{d}t^{4}}
\end{equation}
Using the dot notation for the time derivatives, one gets 
\begin{align}
    DG^\prime_{t}	&=-\begin{vmatrix}\dot{x_{2}} & \dot{x_{3}} & \dot{x_{4}} & \dot{x_{5}}\\
\ddot{x_{2}} & \ddot{x_{3}} & \ddot{x_{4}} & \ddot{x_{5}}\\
\dddot{x_{2}} & \dddot{x_{3}} & \dddot{x_{4}} & \dddot{x_{5}}\\
\ddddot{x_{2}} & \ddddot{x_{3}} & \ddddot{x_{4}} & \ddddot{x_{5}}
\end{vmatrix},\\
DG^{\prime\prime}_{t}	&=\begin{vmatrix}\dot{x_{1}} & \dot{x_{3}} & \dot{x_{4}} & \dot{x_{5}}\\
\ddot{x_{1}} & \ddot{x_{3}} & \ddot{x_{4}} & \ddot{x_{5}}\\
\dddot{x_{1}} & \dddot{x_{3}} & \dddot{x_{4}} & \dddot{x_{5}}\\
\ddddot{x_{1}} & \ddddot{x_{3}} & \ddddot{x_{4}} & \ddddot{x_{5}}
\end{vmatrix},\\
DH^\prime_{t}	&=-\begin{vmatrix}\dot{x_{1}} & \dot{x_{2}} & \dot{x_{4}} & \dot{x_{5}}\\
\ddot{x_{1}} & \ddot{x_{2}} & \ddot{x_{4}} & \ddot{x_{5}}\\
\dddot{x_{1}} & \dddot{x_{2}} & \dddot{x_{4}} & \dddot{x_{5}}\\
\ddddot{x_{1}} & \ddddot{x_{2}} & \ddddot{x_{4}} & \ddddot{x_{5}}
\end{vmatrix},\\
DH^{\prime\prime}_{t}	&=\begin{vmatrix}\dot{x_{1}} & \dot{x_{2}} & \dot{x_{3}} & \dot{x_{5}}\\
\ddot{x_{1}} & \ddot{x_{2}} & \ddot{x_{3}} & \ddot{x_{5}}\\
\dddot{x_{1}} & \dddot{x_{2}} & \dddot{x_{3}} & \dddot{x_{5}}\\
\ddddot{x_{1}} & \ddddot{x_{2}} & \ddddot{x_{3}} & \ddddot{x_{5}}
\end{vmatrix},\\
D	&=\begin{vmatrix}\dot{x_{1}} & \dot{x_{2}} & \dot{x_{4}} & \dot{x_{4}}\\
\ddot{x_{1}} & \ddot{x_{2}} & \ddot{x_{4}} & \ddot{x_{4}}\\
\dddot{x_{1}} & \dddot{x_{2}} & \dddot{x_{4}} & \dddot{x_{4}}\\
\ddddot{x_{1}} & \ddddot{x_{2}} & \ddddot{x_{4}} & \ddddot{x_{4}}
\end{vmatrix}.
\end{align}

It is interesting to note that the whole derivation above is also valid in a stress controlled framework where we replace the position vector $\boldsymbol{x}=\left(\varepsilon,\dot{\varepsilon}/\omega,\varepsilon^2,2\dot{\varepsilon}\varepsilon/\omega,S\right)$ by $\boldsymbol{x}=\left(S_{xy},\dot{S}_{xy}/\omega,S_{xx},\dot{S}_{xx}/\omega,\vartheta\right)$ with $\vartheta\in\left\{\varepsilon,\varepsilon^2\right\}$ to get the equivalent of compliances $J_t^{\prime}$, $J_t^{\prime\prime}$ in large transformations that we can call $J_t^{\prime}$, $J_t^{\prime\prime}$, $K_t^{\prime}$ and $K_t^{\prime\prime}$.

\section{Applications}\label{sec:Applications}

In the following section, two classical linear models will be tackled to illustrate the similarities and the differences between the usual Sequence  of Physical Processes and the new extension of it: the Maxwell model\cite{Maxwell1867} and the Kelvin-Voigt model\cite{Kelvin1890,Voigt1890}. 

\subsection{Maxwell model}

Let us consider a Maxwell model 
\begin{equation}
    \boldsymbol{S}+\lambda\dot{\boldsymbol{S}}=2\eta\dot{\boldsymbol{e}}. \label{eq:MaxwellPiola2}
\end{equation}
The equations are then 
\begin{align}
    S_{xx}+\lambda\dot{S}_{xx}	&=0,\\\label{eq:Sxx2}
S_{xy}+\lambda\dot{S}_{xy}	&=\eta\dot{\varepsilon},\\
S_{xz}+\lambda\dot{S}_{xz}	&=0,\\
S_{yy}+\lambda\dot{S}_{yy}	&=2\eta\dot{\varepsilon}\varepsilon,\\
S_{yz}+\lambda\dot{S}_{yz}	&=0,\\
S_{zz}+\lambda\dot{S}_{zz}	&=0.
\end{align}
Assuming that $\boldsymbol{S}\left(0\right)=\boldsymbol{0}$, one solves to get $S_{zz}=S_{xz}=S_{yz}=S_{xx}=0$ and
\begin{align}
    S_{yy}\left(t\right)	&=2\frac{\eta}{\lambda}\mathrm{e}^{-t/\lambda}\int_{0}^{t}\mathrm{e}^{s/\lambda}\dot{\varepsilon}\left(s\right)\varepsilon\left(s\right)\,\mathrm{d}s,\\
S_{xy}\left(t\right)	&=\frac{\eta}{\lambda}\mathrm{e}^{-t/\lambda}\int_{0}^{t}\mathrm{e}^{s/\lambda}\dot{\varepsilon}\left(s\right)\,\mathrm{d}s.
\end{align}
If one assumes additionally that for all $t\in\mathbb{R}_+$, $\varepsilon\left(t\right)=\varepsilon_0\sin\left(\omega t\right)$ with $\omega$ a certain pulsation, one gets
\begin{align}
	S_{yy}\left(t\right)	&=\eta\omega\varepsilon_{0}^{2}\frac{\sin\left(2\omega t\right)-2\lambda\omega\cos\left(2\omega t\right)+2\lambda\omega\mathrm{e}^{-t/\lambda}}{1+\left(2\lambda\omega\right)^{2}},\\
S_{xy}\left(t\right)	&=\eta\omega\varepsilon_{0}\frac{\lambda\omega\sin\left(\omega t\right)+\cos\left(\omega t\right)-\mathrm{e}^{-t/\lambda}}{1+\left(\lambda\omega\right)^{2}}.
\end{align}
What can be interesting is to look at the permanent oscillatory regime when $t\to+\infty$, which reads 
\begin{align}
	S_{yy}\left(t\right)	&=\eta\omega\varepsilon_{0}^{2}\frac{\sin\left(2\omega t\right)-2\lambda\omega\cos\left(2\omega t\right)}{1+\left(2\lambda\omega\right)^{2}},\label{eq:MaxwellSolutionInfty3}\\
S_{xy}\left(t\right)	&=\eta\omega\varepsilon_{0}\frac{\lambda\omega\sin\left(\omega t\right)+\cos\left(\omega t\right)}{1+\left(\lambda\omega\right)^{2}}.\label{eq:MaxwellSolutionInfty4}
\end{align}

It is blatant that $S_{yy}\propto\varepsilon_0^2$ and $S_{xy}\propto\varepsilon_0$,  thus, when $\varepsilon_0\to0$, $S_{yy}$ will become negligible compared to $S_{xy}$, which is the usual case with small oscillatory shear knowing also that, in this limit, $\sigma_{xy}\approx S_{xy}$. Analysing \cref{eq:MaxwellSolutionInfty3,eq:MaxwellSolutionInfty4}, we recover the usual solution for the shear component $S_{xy}$ replacing $\boldsymbol{S}$ by $\boldsymbol{\sigma}$ in \cref{eq:MaxwellPiola2}. However, there is an axial component $S_{yy}$ which  oscillates with a double frequency $2\omega$ compared to the original strain oscillation. 

Now if we come back to the Cauchy stress tensor, one gets
\begin{equation}
    \boldsymbol{\sigma}=\boldsymbol{S}+\varepsilon\begin{pmatrix}2S_{xy}+\varepsilon S_{yy} & S_{yy} & S_{yz}\\
S_{yy} & 0 & 0\\
S_{yz} & 0 & 0
\end{pmatrix}
\end{equation}
which gives, components by components,
\begin{align}
	\sigma_{xx}	&=2\varepsilon S_{xy}+\varepsilon^{2}S_{yy},\label{eq:CauchyPiolaxx}\\
	\sigma_{xy}	&=S_{xy}+\varepsilon S_{yy},\label{eq:CauchyPiolaxy}\\
	\sigma_{yy}	&=S_{yy},\label{eq:CauchyPiolayy}\\
	\sigma_{xz}	&=\sigma_{yz}=\sigma_{zz}=0.\label{eq:CauchyPiolayz}
\end{align}
Replacing now with \cref{eq:MaxwellSolutionInfty3,eq:MaxwellSolutionInfty4} and doing some trigonometric calculations, one obtains

\begin{align}\sigma_{xx}\left(t\right) & =\eta\omega\varepsilon_{0}^{2}\left(c_{0}^{xx}+c_{2}^{xx}\cos\left(2\omega t\right)+s_{2}^{xx}\sin\left(2\omega t\right)+\right.\\
 & \left.+c_{4}^{xx}\cos\left(4\omega t\right)+s_{4}^{xx}\sin\left(4\omega t\right)\right),\\
c_{0}^{xx} & =\frac{\lambda\omega}{2}\left(\frac{2}{1+\left(\lambda\omega\right)^{2}}+\frac{\varepsilon_{0}^{2}}{1+\left(2\lambda\omega\right)^{2}}\right),\\
c_{2}^{xx} & =-\lambda\omega\left(\frac{1}{1+\left(\lambda\omega\right)^{2}}+\frac{\varepsilon_{0}^{2}}{1+\left(2\lambda\omega\right)^{2}}\right),\\
s_{2}^{xx} & =\frac{1}{2}\left(\frac{2}{1+\left(\lambda\omega\right)^{2}}+\frac{\varepsilon_{0}^{2}}{1+\left(2\lambda\omega\right)^{2}}\right),\\
c_{4}^{xx} & =\frac{\lambda\omega\varepsilon_{0}^{2}}{2}\frac{1}{1+\left(2\lambda\omega\right)^{2}},\\
s_{4}^{xx} & =-\frac{\varepsilon_{0}^{2}}{4}\frac{1}{1+\left(2\lambda\omega\right)^{2}},\\
\sigma_{xy}\left(t\right) & =\eta\omega\varepsilon_{0}\left(c_{1}^{xy}\cos\left(\omega t\right)+s_{1}^{xy}\sin\left(\omega t\right)+\right.\\
 & \left.+c_{3}^{xy}\cos\left(3\omega t\right)+s_{3}^{xy}\sin\left(3\omega t\right)\right),\\
c_{1}^{xy} & =\frac{1}{1+\left(\lambda\omega\right)^{2}}+\frac{\varepsilon_{0}}{2}\frac{1}{1+\left(2\lambda\omega\right)^{2}},\\
s_{1}^{xy} & =\lambda\omega\left(\frac{1}{1+\left(\lambda\omega\right)^{2}}+\frac{\varepsilon_{0}}{1+\left(2\lambda\omega\right)^{2}}\right),\\
c_{3}^{xy} & =-\frac{\varepsilon_{0}}{2}\frac{1}{1+\left(2\lambda\omega\right)^{2}},\\
s_{3}^{xy} & =-\frac{\lambda\omega\varepsilon_{0}}{1+\left(2\lambda\omega\right)^{2}},\\
\sigma_{yy}\left(t\right) & =\eta\omega\varepsilon_{0}^{2}\frac{\sin\left(2\omega t\right)-2\lambda\omega\cos\left(2\omega t\right)}{1+\left(2\lambda\omega\right)^{2}}.
\end{align}

What is really interesting from the equations above is that, in the Cauchy stress tensor, there are non zero $\sigma_{xy}$ and $\sigma_{yy}$ components but also the $\sigma_{xx}$ component. Also, the $\sigma_{yy}$ component remains identical to $S_{yy}$, with the double frequency oscillation, but $\sigma_{xy}$ has two harmonics, the first and the third, and $\sigma_{xx}$ has three harmonics, the zeroth, the second and the fourth. The non-zero average of $\sigma_{xx}$ is equal to \begin{multline}
	\lim_{T\to+\infty}\frac{1}{T}\int_{0}^{T}\sigma_{xx}\left(t\right)\,\mathrm{d}t=\\\frac{\eta\lambda\omega^2\varepsilon_{0}^{2}}{2}\left(\frac{2}{1+\left(\lambda\omega\right)^{2}}+\frac{\varepsilon_{0}^{2}}{1+\left(2\lambda\omega\right)^{2}}\right).
\end{multline}

It is now possible to compare the usual Sequence of Physical Process framework with the extended version presented here. In the usual Sequence of Physical Process, the transient moduli are calculated through  
\begin{align}
	\mathcal{G}_{t}^{\prime}\left(\sigma_{xy}\right)	&=-\frac{\ddot{\varepsilon}\ddot{\sigma}_{xy}-\dddot{\varepsilon}\dot{\sigma}_{xy}}{\dot{\varepsilon}\dddot{\varepsilon}-\ddot{\varepsilon}^{2}},\label{eq:UsualGp}\\
	\mathcal{G}_{t}^{\prime\prime}\left(\sigma_{xy}\right)	&=-\omega\frac{\dot{\sigma}_{xy}\ddot{\varepsilon}-\ddot{\sigma}_{xy}\dot{\varepsilon}}{\dot{\varepsilon}\dddot{\varepsilon}-\ddot{\varepsilon}^{2}}\label{eq:UsualGpp}.
\end{align}
Using the expressions above in the limit $\varepsilon_0\to0$, one obtains
\begin{align}
	\mathcal{G}_{t}^{\prime}\left(\sigma_{xy}\right)	&=\eta\omega\frac{\lambda\omega}{1+\left(\lambda\omega\right)^{2}},\label{eq:ClassicalSPP1}\\
	\mathcal{G}_{t}^{\prime\prime}\left(\sigma_{xy}\right)	&=\frac{\eta\omega}{1+\left(\lambda\omega\right)^{2}}\label{eq:ClassicalSPP2}
\end{align}
as a usual linear Maxwell model with the variables $\boldsymbol{\sigma}$ and $\dot{\boldsymbol{\varepsilon}}$. Using the new extended version with the second Piola-Kirchhoff tensor, it is possible to obtain with the previous notations 
\begin{equation}
	D=\begin{vmatrix}\dot{x}_{1} & \dot{x}_{2} & \dot{x}_{3} & \dot{x}_{4}\\
\ddot{x}_{1} & \ddot{x}_{2} & \ddot{x}_{3} & \ddot{x}_{4}\\
\dddot{x}_{1} & \dddot{x}_{2} & \dddot{x}_{3} & \dddot{x}_{4}\\
\ddddot{x}_{1} & \ddddot{x}_{2} & \ddddot{x}_{3} & \ddddot{x}_{4}
\end{vmatrix}=18\varepsilon_{0}^{6}\omega^{10}
\end{equation}
then doing the calculations for $S_{xy}$ brings
\begin{align}
	G_{t}^{\prime}\left(S_{xy}\right)	&=\eta\omega\frac{\lambda\omega}{1+\left(\lambda\omega\right)^{2}},\label{eq:GtpSxy}\\
G_{t}^{\prime\prime}\left(S_{xy}\right)	&=\frac{\eta\omega}{1+\left(\lambda\omega\right)^{2}}\label{eq:GtppSxy},\\
H_{t}^{\prime}\left(S_{xy}\right)&=0,\\
H_{t}^{\prime\prime}\left(S_{xy}\right)&=0,
\end{align}
and for $S_{yy}$ brings
\begin{align}
	G_{t}^{\prime}\left(S_{yy}\right)	&=0,\\
G_{t}^{\prime\prime}\left(S_{yy}\right)	&=0,\\
H_{t}^{\prime}\left(S_{yy}\right)&=\eta\omega\frac{4\lambda\omega}{1+\left(2\lambda\omega\right)^{2}},\label{eq:HtpSxx}\\
H_{t}^{\prime\prime}\left(S_{yy}\right)&=\frac{2\eta\omega}{1+\left(2\lambda\omega\right)^{2}}.\label{eq:HtppSxx}
\end{align}

Hence, we find transient moduli both in the shear direction $xy$ and in the axial direction $xx$ with very interesting relationships like 
\begin{equation}
	\frac{H_{t}^{\prime}\left(S_{yy}\right)}{H_{t}^{\prime\prime}\left(S_{yy}\right)}=2\frac{G_{t}^{\prime}\left(S_{xy}\right)}{G_{t}^{\prime\prime}\left(S_{xy}\right)}=2\lambda\omega.
\end{equation}

Another interesting feature which may be highlighted is the fact that, due to construction with \cref{eq:LinearDecomposition} and looking at \cref{eq:MaxwellSolutionInfty3,eq:MaxwellSolutionInfty4,eq:GtpSxy,eq:GtppSxy,eq:HtpSxx,eq:HtppSxx}, each component of $\boldsymbol{S}$ is a linear combination of $\varepsilon$, $\dot{\varepsilon}/\omega$, $\varepsilon^2$ and $2\dot{\varepsilon}\varepsilon/\omega$ with the transient moduli $G_t^{\prime}$, $G_t^{\prime\prime}$, $H_t^{\prime}$ and $H_t^{\prime\prime}$ as factors.

It is now possible to find the last moduli for $\sigma_{xx}$, $\sigma_{xy}$ and $\sigma_{yy}$. The easiest one is $\sigma_{yy}$ thanks to \cref{eq:CauchyPiolayy} thus  
\begin{align}
	G_{t}^{\prime}\left(\sigma_{yy}\right)	&=0,\\
G_{t}^{\prime\prime}\left(\sigma_{yy}\right)	&=0,\\
H_{t}^{\prime}\left(\sigma_{yy}\right)&=\eta\omega\frac{4\lambda\omega}{1+\left(2\lambda\omega\right)^{2}},\\
H_{t}^{\prime\prime}\left(\sigma_{yy}\right)&=\frac{2\eta\omega}{1+\left(2\lambda\omega\right)^{2}}.
\end{align}
For $\sigma_{xx}$ and $\sigma_{xy}$, the fourth and the third harmonics, respectively, make calculations more tedious so the overall framework should be applied. In the case of $\sigma_{xx}$, one gets  
\begin{align}
	G_{t}^{\prime}\left(\sigma_{xx}\right)	&=8\eta\omega\varepsilon_0\left(c_4^{xx}\left(3\sin\left(5\omega t\right)-5\sin\left(3\omega t\right)\right)+\right.\\
	&\left.+s_4^{xx}\left(5\cos\left(3\omega t\right)-3\cos\left(5\omega t\right)\right)\right)\\
G_{t}^{\prime\prime}\left(\sigma_{xx}\right)	&=8\eta\omega\varepsilon_0\left(c_4^{xx}\left(3\cos\left(5\omega t\right)+5\cos\left(3\omega t\right)\right)+\right.\\
	&\left.+s_4^{xx}\left(5\sin\left(3\omega t\right)+3\sin\left(5\omega t\right)\right)\right)\\
H_{t}^{\prime}\left(\sigma_{xx}\right)&=-2\eta\omega\left(c_2^{xx}+\right.\\
 &\left.+5c_4^{xx}\left(\cos\left(6\omega t\right)+3\cos\left(2\omega t\right)\right)+\right.\\
 &\left.+5s_4^{xx}\left(\sin\left(6\omega t\right)+3\sin\left(2\omega t\right)\right)\right)\\
H_{t}^{\prime\prime}\left(\sigma_{xx}\right)&=2\eta\omega\left(s_2^{xx}+\right.\\
 &\left.+5c_4^{xx}\left(\sin\left(6\omega t\right)-3\sin\left(2\omega t\right)\right)+\right.\\
 &\left.+5s_4^{xx}\left(-\cos\left(6\omega t\right)+3\cos\left(2\omega t\right)\right)\right)
\end{align} 
and for $\sigma_{xy}$, one obtains
\begin{align}
	G_{t}^{\prime}\left(\sigma_{xy}\right)	&=\eta\omega\left(s_1^{xy}+\right.\label{eq:ExtendedSPP1}\\
	&\left.+5c_3^{xy}\left(2\sin\left(2\omega t\right)-\sin\left(4\omega t\right)\right)\right.\\
	&\left.+5s_3^{xy}\left(\cos\left(4\omega t\right)-2\cos\left(2\omega t\right)\right)\right)\\
G_{t}^{\prime\prime}\left(\sigma_{xy}\right)	&=\eta\omega\left(c_1^{xy}+\right.\\
	&\left.-5c_3^{xy}\left(2\cos\left(2\omega t\right)+\cos\left(4\omega t\right)\right)\right.\\
	&\left.-5s_3^{xy}\left(\sin\left(4\omega t\right)+2\sin\left(2\omega t\right)\right)\right)\label{eq:ExtendedSPP2}\\
H_{t}^{\prime}\left(\sigma_{xy}\right)&=\frac{2\eta\omega}{\varepsilon_0}\left(c_3^{xy}\left(5\cos\left(\omega t\right)+\cos\left(5\omega t\right)\right)+\right.\\
&\left.+s_3^{xy}\left(5\sin\left(\omega t\right)+\sin\left(5\omega t\right)\right)\right)\\
H_{t}^{\prime\prime}\left(\sigma_{xy}\right)&=\frac{2\eta\omega}{\varepsilon_0}\left(c_3^{xy}\left(-5\sin\left(\omega t\right)+\sin\left(5\omega t\right)\right)+\right.\\
&\left.+s_3^{xy}\left(5\cos\left(\omega t\right)-\cos\left(5\omega t\right)\right)\right)
\end{align} 

With the previous equations, it can be interesting to compare the usual Sequence of Physical Processes with the present extended version. Thus, using \cref{eq:ClassicalSPP1,eq:ClassicalSPP2} and the average values according to time of \crefrange{eq:ExtendedSPP1}{eq:ExtendedSPP2}, one obtains
\begin{align}
	\left\langle G_{t}^{\prime}\left(\sigma_{xy}\right)\right\rangle-\mathcal{G}_{t}^{\prime}\left(\sigma_{xy}\right)&=\eta\omega\varepsilon_0\frac{\lambda\omega}{1+\left(2\lambda\omega\right)^2}\\
	\left\langle G_{t}^{\prime\prime}\left(\sigma_{xy}\right)\right\rangle-\mathcal{G}_{t}^{\prime\prime}\left(\sigma_{xy}\right)&=\frac{\eta\omega\varepsilon_0}{2}\frac{1}{1+\left(2\lambda\omega\right)^2}.
\end{align}

\begin{figure*}
	\centering
	\includegraphics[width=\textwidth]{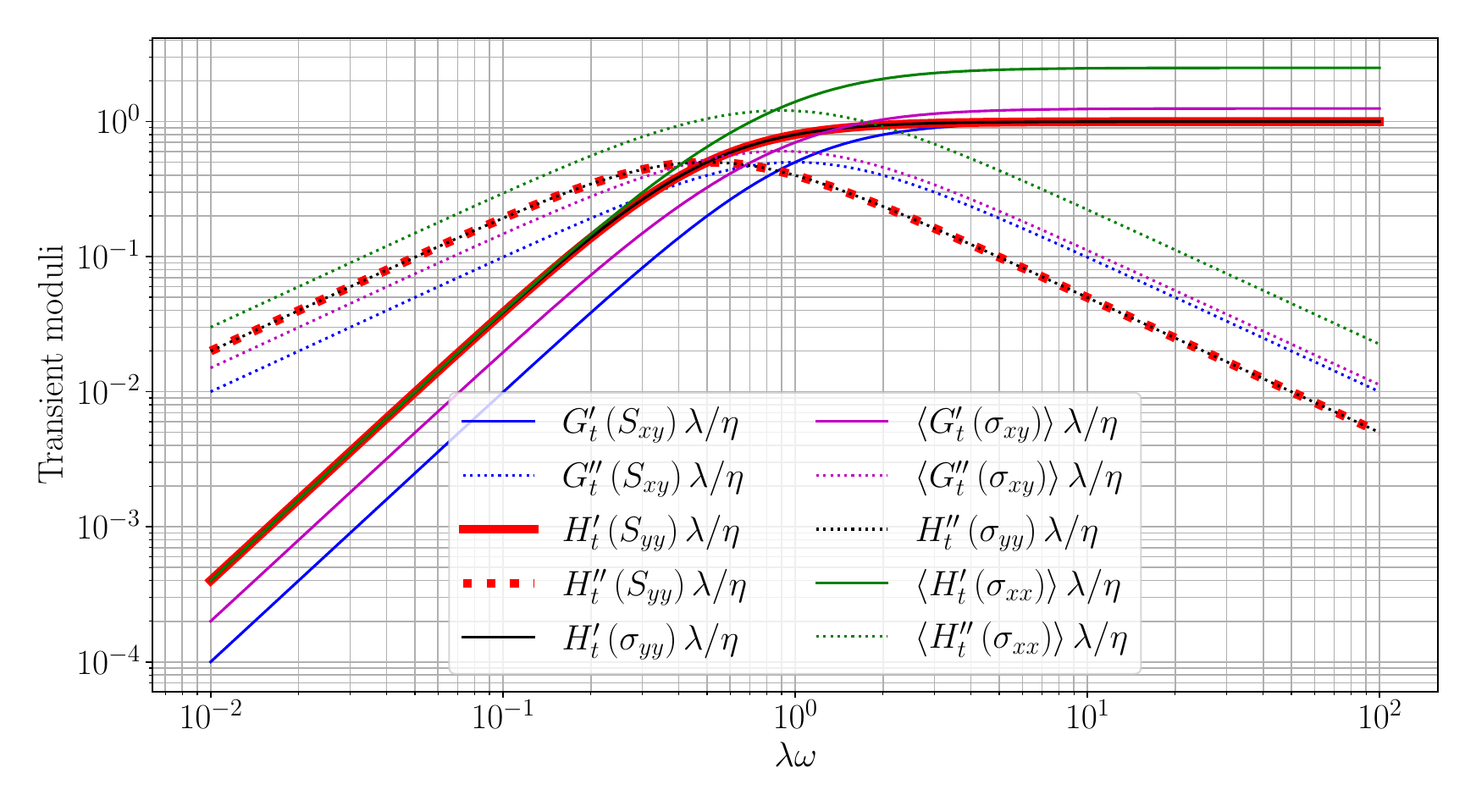}
	\caption{Evolution of the transient moduli of $S_{xy}$, $S_{yy}$ and $\sigma_{yy}$ and the average according to time of transient moduli of $\sigma_{xx}$ and $\sigma_{xy}$, all divided by $\eta/\lambda$ as a function $\lambda\omega$ with $\varepsilon_0=1$.}
	\label{fig:FrequencyModuliPiola}
\end{figure*}

To give more clarity of understanding, it is possible to plot the variations of these moduli according to different parameters. Noting that the moduli of $S_{xy}$, $S_{yy}$ and $\sigma_{yy}$ are constant with time, \cref{fig:FrequencyModuliPiola} illustrates the frequency plot of those moduli. We recognise easily the usual behaviour of the linear Maxwell model for $S_{xy}$ but the contribution on $S_{yy}$ or $\sigma_{yy}$ is of the same order of magnitude when $\varepsilon_0$ is of the order of unity or higher. Also, the time scales are a bit different and shows an earlier shift in the trends.   

\begin{figure*}
	\centering
	\includegraphics[width=\textwidth]{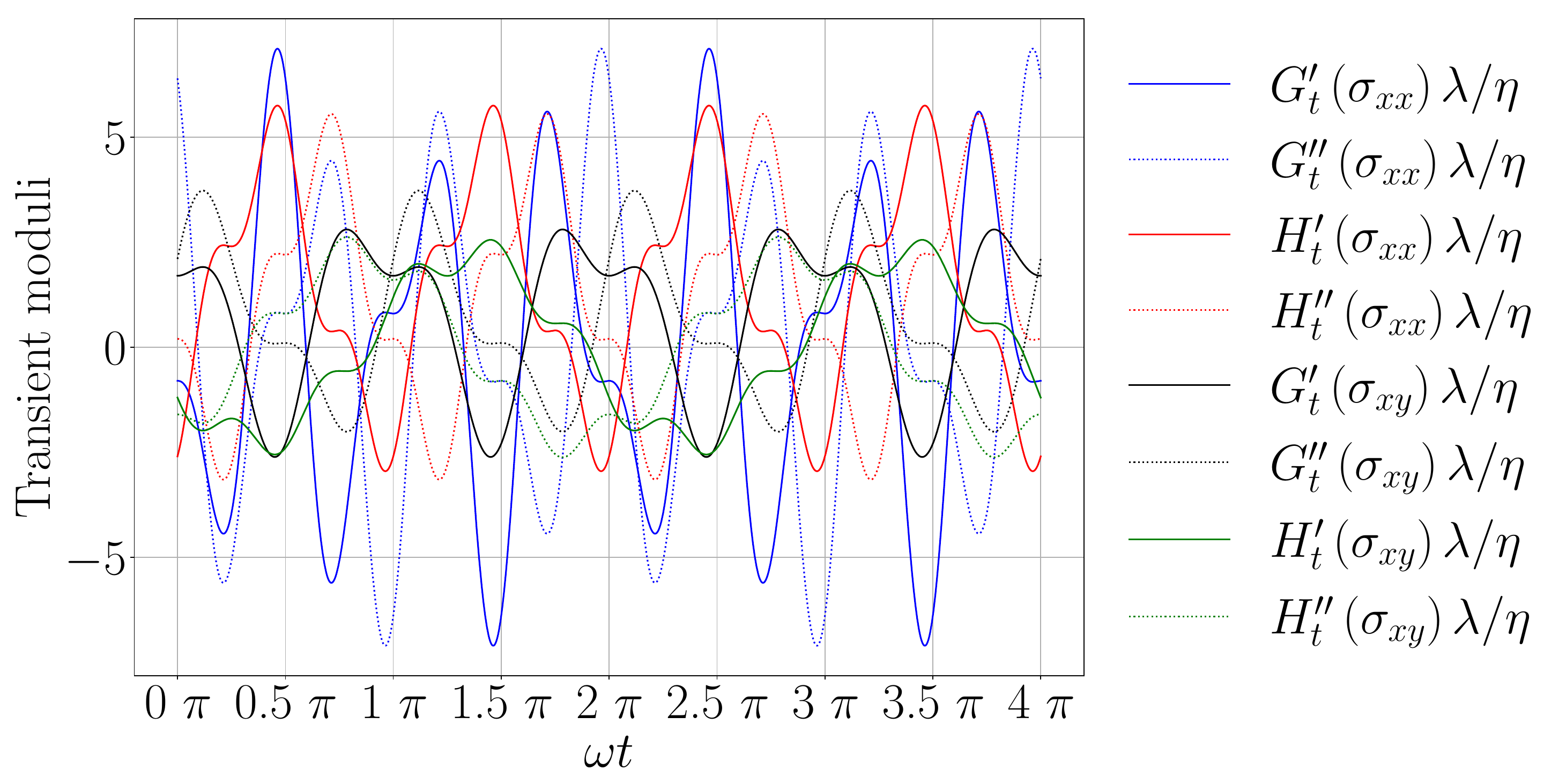}
	\caption{Evolution of the various moduli of $\sigma_{xy}$ and $\sigma_{xx}$ divided by $\eta/\lambda$ as a function $\omega t$ imposing $\lambda\omega=\varepsilon_0=1$.}
	\label{fig:TimeModuliPiola}
\end{figure*}

\begin{figure}
	\centering
	\includegraphics[width=\columnwidth]{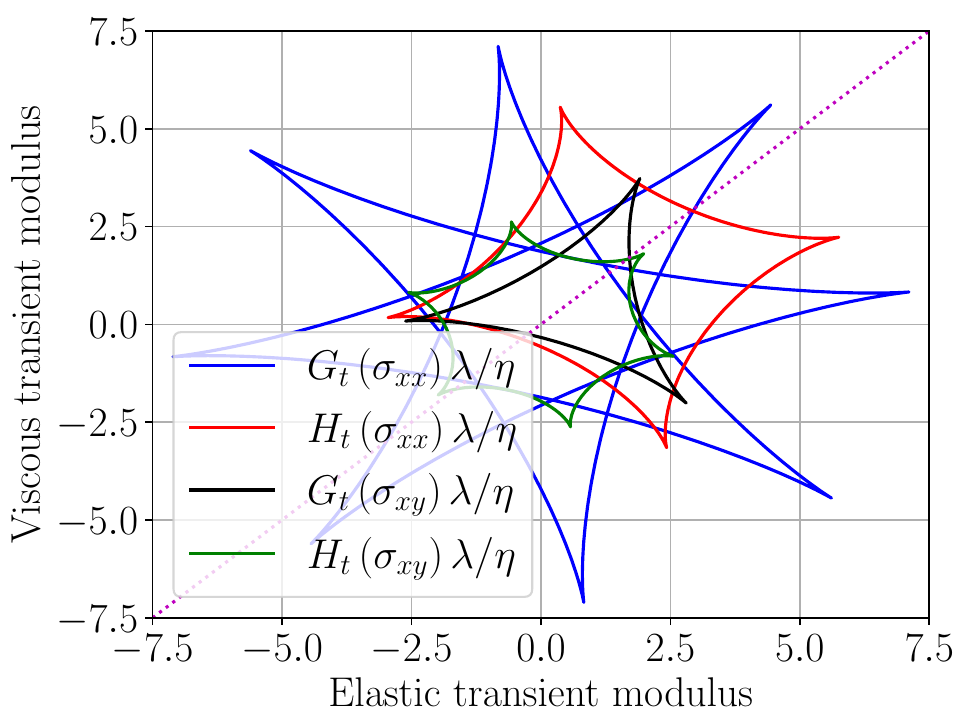}
	\caption{Evolution of the viscous moduli of $\sigma_{xy}$ and $\sigma_{xx}$ divided by $\eta/\lambda$ as a function of the elastic moduli divided by $\eta/\lambda$ imposing $\lambda\omega=\varepsilon_0=1$. The purple dotted line corresponds to the first bisector $y=x$.}
	\label{fig:PhaseModuliPiola}
\end{figure}

Now if we look at the expressions for $\sigma_{xx}$ and $\sigma_{xy}$, we see that there are non zero contribution of each component. Normally, one would expect only the $\sigma_{xy}$ component to play a role but here, due to the increase of the strain amplitude $\varepsilon_0$, additional contributions appear. It is possible to recover the usual linear Maxwell model expressed in terms of the Cauchy stress tensor $\boldsymbol{\sigma}$ when $\varepsilon_0\to0$. However, at moderately high strain amplitude, there are variations in time of the various transient moduli. There is the actual novelty of this approach : with the steady-state approach, one would only find the average values over time. Now, one can find some more intricate evolutions to properly understand the materials behaviour. To give some illustrations to those equations, \cref{fig:TimeModuliPiola} shows the time evolution of the various transient moduli of $\sigma_{xx}$ and $\sigma_{xy}$ with the parameters $\varepsilon_0=\lambda\omega=1$. To give another perspective, \cref{fig:PhaseModuliPiola} gives the Cole-Cole plot, i.e. the viscous transient moduli as a function of the elastic transient moduli, with the same set of parameters. An interesting thing when looking at \cref{fig:PhaseModuliPiola} is that, focusing on the black line for $G_t\left(\sigma_{xy}\right)$, which is usually observed in experiments presenting the 3 peaks, there exist some portion of the cycle where either or both $G_t^{\prime}\left(\sigma_{xy}\right)$ and $G_t^{\prime\prime}\left(\sigma_{xy}\right)$ are negative. This is interpreted sometimes as sign of elastic recoil or negative dissipation. However, this example demonstrates that even for a model where the rheological parameters are properly defined without any doubts on thermodynamics, there is some strange behaviour happening which may lead to skewed interpretations. In this case, a simple geometrical transformation of the Cauchy stress tensor $\boldsymbol{\sigma}$ into the second Piola-Kirchhoff stress tensor $\boldsymbol{S}$ leads to a rheological analysis which is much simpler due to the fact that the newly obtained moduli will be constant and corresponding exactly to a linear Maxwell fluid. All these remarks are just a warning for people who are looking for a characterisation and parameter identification of their materials: some processing may be needed on experimental data to analyse properly the rheological behaviour having taken into account geometrical non linearities.

\subsection{Kelvin-Voigt model}

Let us consider a Kelvin-Voigt model 
\begin{equation}
    \boldsymbol{S}=2E\boldsymbol{e}+2\eta\dot{\boldsymbol{e}}. \label{eq:KelvinPiola}
\end{equation}
The equations are then 
\begin{align}
    S_{xx}	&=0,\\
    S_{xy}  &=E\varepsilon+\eta\dot{\varepsilon},\\
    S_{xz}	&=0,\\
    S_{yy}  &= E\varepsilon^2+2\eta\varepsilon\dot{\varepsilon},\\
    S_{yz}	&=0,\\
    S_{zz}	&=0.
\end{align}
If one assumes additionally that for all $t\in\mathbb{R}_+$, $\varepsilon\left(t\right)=\varepsilon_0\sin\left(\omega t\right)$ with $\omega$ a certain pulsation, one gets
\begin{align}
	S_{yy}\left(t\right)	&=E\varepsilon_0^2\left(1-\cos\left(2\omega t\right)\right)+\eta\omega\varepsilon_0^2\sin\left(2\omega t\right),\label{eq:KelvinSolution1}\\
    S_{xy}\left(t\right)	&=E\varepsilon_0\sin\left(\omega t\right)+\eta\omega\varepsilon_0\cos\left(\omega t\right)\label{eq:KelvinSolution2}.
\end{align}

It is again blatant that $S_{yy}\propto\varepsilon_0^2$ and $S_{xy}\propto\varepsilon_0$,  thus, when $\varepsilon_0\to0$, $S_{yy}$ will become negligible compared to $S_{xy}$, which is the usual case with small oscillatory shear knowing also that, in this limit, $\sigma_{xy}\approx S_{xy}$. Analysing \cref{eq:KelvinSolution1,eq:KelvinSolution2}, we recover the usual solution for the shear component $S_{xy}$ replacing $\boldsymbol{S}$ by $\boldsymbol{\sigma}$ in \cref{eq:KelvinPiola} ; however, there is an axial component $S_{yy}$ which  oscillates with a double frequency $2\omega$ compared to the original strain oscillation. Another interesting fact is that $S_{yy}$ has a non-zero zeroth harmonic which is equal to $E\varepsilon_0^2$.

Now if we come back to the Cauchy stress tensor, one gets with \cref{eq:CauchyPiolaxx,eq:CauchyPiolaxy,eq:CauchyPiolayy,eq:CauchyPiolayz,eq:KelvinSolution1,eq:KelvinSolution2} doing some trigonometric calculations,

\begin{align}
\sigma_{xx}\left(t\right) & =E\varepsilon_{0}^{2}\left(1+\frac{3}{4}\varepsilon_{0}^{2}\right)-E\varepsilon_{0}^{2}\cos\left(2\omega t\right)\left(1+\varepsilon_{0}^{2}\right)\\
    &+\eta\omega\varepsilon_{0}^{2}\left(1+\frac{\varepsilon_{0}^{2}}{2}\right)\sin\left(2\omega t\right)+\frac{E\varepsilon_{0}^{4}}{4}\cos\left(4\omega t\right)\\
    &-\frac{\eta\omega\varepsilon_{0}^{4}}{4}\sin\left(4\omega t\right),\\
\sigma_{xy}\left(t\right) & =E\varepsilon_{0}\left(1+\frac{3}{2}\varepsilon_{0}^{2}\right)\sin\left(\omega t\right)\\
    &+\eta\omega\varepsilon_{0}\left(1+\frac{\varepsilon_{0}^{2}}{2}\right)\cos\left(\omega t\right)-\frac{E\varepsilon_{0}^{3}}{2}\sin\left(3\omega t\right)\\
    &-\frac{\eta\omega\varepsilon_{0}^{3}}{2}\cos\left(3\omega t\right),\\
\sigma_{yy}\left(t\right) & =E\varepsilon_0^2\left(1-\cos\left(2\omega t\right)\right)+\eta\omega\varepsilon_0^2\sin\left(2\omega t\right).
\end{align}

What is really noteworthy from the equations above is that, in the Cauchy stress tensor, there are non zero $\sigma_{xy}$ and $\sigma_{yy}$ components but also a $\sigma_{xx}$ component. Also, the $\sigma_{yy}$ component remains identical to $S_{yy}$, with the double frequency oscillation, but $\sigma_{xy}$ has two harmonics, the first and the third, and $\sigma_{xx}$ has three harmonics, the zeroth, the second and the fourth. The non-zero average of $\sigma_{xx}$ is equal to 
\begin{equation}
	\lim_{T\to+\infty}\frac{1}{T}\int_{0}^{T}\sigma_{xx}\left(t\right)\,\mathrm{d}t=E\varepsilon_{0}^{2}\left(1+\frac{3}{4}\varepsilon_{0}^{2}\right).
\end{equation}

It is now possible to compare the usual Sequence of Physical Process with the extended version presented here. In the usual Sequence of Physical Process, the transient moduli are calculated through \cref{eq:UsualGp,eq:UsualGpp} which gives, using the expressions above in the limit $\varepsilon_0\to0$, 

\begin{align}
	\mathcal{G}_{t}^{\prime}\left(\sigma_{xy}\right)	&=E\label{eq:ClassicalSPP1Kelvin}\\
	\mathcal{G}_{t}^{\prime\prime}\left(\sigma_{xy}\right)	&=\eta\omega\label{eq:ClassicalSPP2Kelvin}
\end{align}
 as a usual linear Kelvin-Voigt model with the variables $\boldsymbol{\sigma}$ and $\boldsymbol{\varepsilon}$. Using the new extended version with the second Piola-Kirchhoff tensor, it is possible to obtain for $S_{xy}$,
\begin{align}
	G_{t}^{\prime}\left(S_{xy}\right)	&=E,\label{eq:GtpSxyKelvin}\\
G_{t}^{\prime\prime}\left(S_{xy}\right)	&=\eta\omega\label{eq:GtppSxyKelvin},\\
H_{t}^{\prime}\left(S_{xy}\right)&=0,\\
H_{t}^{\prime\prime}\left(S_{xy}\right)&=0,
\end{align}
and for $S_{yy}$,
\begin{align}
	G_{t}^{\prime}\left(S_{yy}\right)	&=0,\\
G_{t}^{\prime\prime}\left(S_{yy}\right)	&=0,\\
H_{t}^{\prime}\left(S_{yy}\right)&=E,\label{eq:HtpSxxKelvin}\\
H_{t}^{\prime\prime}\left(S_{yy}\right)&=\eta\omega.\label{eq:HtppSxxKelvin}
\end{align}

Hence, we find transient moduli both in the shear direction $xy$ and in the axial direction $yy$ with very interesting relationships like 
\begin{align}
	H_{t}^{\prime}\left(S_{yy}\right)=G_{t}^{\prime}\left(S_{xy}\right),\\
    H_{t}^{\prime\prime}\left(S_{yy}\right)=G_{t}^{\prime\prime}\left(S_{xy}\right).
\end{align}

Another interesting feature which may be highlighted is the fact that, due to construction with \cref{eq:LinearDecomposition} and looking at \cref{eq:KelvinSolution1,eq:KelvinSolution2,eq:GtpSxyKelvin,eq:GtppSxyKelvin,eq:HtpSxxKelvin,eq:HtppSxxKelvin}, each component of $\boldsymbol{S}$ is a linear combination of $\varepsilon$, $\dot{\varepsilon}/\omega$, $\varepsilon^2$ and $2\dot{\varepsilon}\varepsilon/\omega$ with the transient moduli $G_t^{\prime}$, $G_t^{\prime\prime}$, $H_t^{\prime}$ and $H_t^{\prime\prime}$ as factors.

It is now possible to find the last moduli for $\sigma_{xx}$, $\sigma_{xy}$ and $\sigma_{yy}$. The easiest one is $\sigma_{yy}$ thanks to \cref{eq:CauchyPiolayy} thus  
\begin{align}
	G_{t}^{\prime}\left(\sigma_{yy}\right)	&=0,\\
G_{t}^{\prime\prime}\left(\sigma_{yy}\right)	&=0,\\
H_{t}^{\prime}\left(\sigma_{yy}\right)&=E,\label{eq:HtpsyyKelvin}\\
H_{t}^{\prime\prime}\left(\sigma_{yy}\right)&=\eta\omega.\label{eq:HtppsyyKelvin}
\end{align}
For $\sigma_{xx}$ and $\sigma_{xy}$, the fourth and the third harmonics, respectively, prevent a quick calculations so the overall framework should be applied. In the case of $\sigma_{xx}$, defining $\tan\left(\alpha\right)=\eta\omega/E$, one gets  
\begin{align}
	\frac{G_{t}^{\prime}\left(\sigma_{xx}\right)}{2\varepsilon_0^3\sqrt{E^2+\eta^2\omega^2}}	&=3\sin\left(5\omega t+\alpha\right)-5\sin\left(3\omega t+\alpha\right)\\
\frac{G_{t}^{\prime\prime}\left(\sigma_{xx}\right)}{2\varepsilon_0^3\sqrt{E^2+\eta^2\omega^2}}	&=3\cos\left(5\omega t+\alpha\right)+5\cos\left(3\omega t+\alpha\right)\\
\frac{H_{t}^{\prime}\left(\sigma_{xx}\right)}{2\sqrt{E^2+\eta^2\omega^2}}&=\cos\left(\alpha\right)\left(1+\varepsilon_{0}^{2}\right)+\label{eq:HtpsxxKelvin}\\
 &-\frac{5\varepsilon_{0}^{2}}{4}\left(3\cos\left(2\omega t+\alpha\right)+\cos\left(6\omega t+\alpha\right)\right)\\
\frac{H_{t}^{\prime\prime}\left(\sigma_{xx}\right)}{2\sqrt{E^2+\eta^2\omega^2}}&=\sin\left(\alpha\right)\left(1+\frac{\varepsilon_{0}^{2}}{2}\right)+\label{eq:HtppsxxKelvin}\\
 &+\frac{5\varepsilon_{0}^{2}}{4}\left(\sin\left(6\omega t+\alpha\right)-3\sin\left(2\omega t+\alpha\right)\right)
\end{align} 
and for $\sigma_{xy}$, one obtains
\begin{align}
	\frac{G_{t}^{\prime}\left(\sigma_{xy}\right)}{\sqrt{E^2+\eta^2\omega^2}}	&=\cos\left(\alpha\right)\left(1+\frac{3}{2}\varepsilon_{0}^{2}\right)+\label{eq:ExtendedSPP1Kelvin}\\
	&-\frac{5\varepsilon_{0}^{2}}{2}\left(\cos\left(4\omega t+\alpha\right)-2\cos\left(2\omega t+\alpha\right)\right)\\
\frac{G_{t}^{\prime\prime}\left(\sigma_{xy}\right)}{\sqrt{E^2+\eta^2\omega^2}}	&=\sin\left(\alpha\right)\left(1+\frac{\varepsilon_{0}^{2}}{2}\right)+\label{eq:GtppsxyKelvin}\\
	&+\frac{5\varepsilon_{0}^{2}}{2}\left(\sin\left(4\omega t+\alpha\right)+2\sin\left(2\omega t+\alpha\right)\right)	\label{eq:ExtendedSPP2Kelvin}\\
\frac{H_{t}^{\prime}\left(\sigma_{xy}\right)}{\varepsilon_0\sqrt{E^2+\eta^2\omega^2}}&=-5\sin\left(\omega t+\alpha\right)-\sin\left(5\omega t+\alpha\right)\\
\frac{H_{t}^{\prime\prime}\left(\sigma_{xy}\right)}{\varepsilon_0\sqrt{E^2+\eta^2\omega^2}}&=\cos\left(5\omega t+\alpha\right)-5\cos\left(\omega t+\alpha\right)
\end{align} 

With the previous equations, it can be interesting to compare the usual Sequence of Physical Processes with the present extended version. Thus, using \cref{eq:ClassicalSPP1Kelvin,eq:ClassicalSPP2Kelvin} and the average time values of \crefrange{eq:ExtendedSPP1Kelvin}{eq:ExtendedSPP2Kelvin}, one obtains
\begin{align}
	\left\langle G_{t}^{\prime}\left(\sigma_{xy}\right)\right\rangle-\mathcal{G}_{t}^{\prime}\left(\sigma_{xy}\right)&=\frac{3}{2}E\varepsilon_{0}^{2}\\
	\left\langle G_{t}^{\prime\prime}\left(\sigma_{xy}\right)\right\rangle-\mathcal{G}_{t}^{\prime\prime}\left(\sigma_{xy}\right)&=\frac{\eta\omega\varepsilon_{0}^{2}}{2}.
\end{align}

To give more clarity of understanding, it is possible to plot the variations of these moduli according to different parameters. Noting that the moduli of $S_{xy}$, $S_{yy}$ and $\sigma_{yy}$ are constant with time, the \cref{tab:KelvinTable} illustrates the summary of the expressions of those moduli. We recognise the usual Kelvin-Voigt model expressions with some additional items due to the geometrical complements of the Piola-Kirchhoff expressions.   

\begin{table}[]
    \centering
    \begin{tabular}{ccc}
        \hline
        Transient moduli & $\bullet_{t}^{\prime}$ & $\bullet_{t}^{\prime\prime}$ \\
        \hline
        $G_{t}\left(S_{xy}\right)$ & $E$ & $\eta\omega$\\
        $H_{t}\left(S_{yy}\right)$ & $E$ & $\eta\omega$\\
        $H_{t}\left(\sigma_{yy}\right)$ & $E$ & $\eta\omega$\\
        $\left\langle H_{t}\left(\sigma_{xx}\right)\right\rangle$ & $2E\left(1+\varepsilon_{0}^{2}\right)$ & $2\eta\omega\left(1+\dfrac{\varepsilon_{0}^{2}}{2}\right)$\\
        $\left\langle G_{t}\left(\sigma_{xy}\right)\right\rangle$ & $E\left(1+\dfrac{3}{2}\varepsilon_{0}^{2}\right)$ & $\eta\omega\left(1+\dfrac{\varepsilon_{0}^{2}}{2}\right)$ \\
        \hline
    \end{tabular}
    \caption{Summary of the expressions of the transient moduli for $S_{xy}$, $S_{yy}$ and $\sigma_{yy}$ and the average overtime of the transient moduli for $\sigma_{xx}$ and $\sigma_{xy}$ according to \cref{eq:GtpSxyKelvin,eq:GtppSxyKelvin,eq:HtppSxxKelvin,eq:HtpSxxKelvin,eq:HtppsyyKelvin,eq:HtpsyyKelvin,eq:HtpsxxKelvin,eq:HtppsxxKelvin,eq:ExtendedSPP1Kelvin,eq:GtppsxyKelvin}.}
    \label{tab:KelvinTable}
\end{table}

\begin{figure*}
	\centering
	\includegraphics[width=\textwidth]{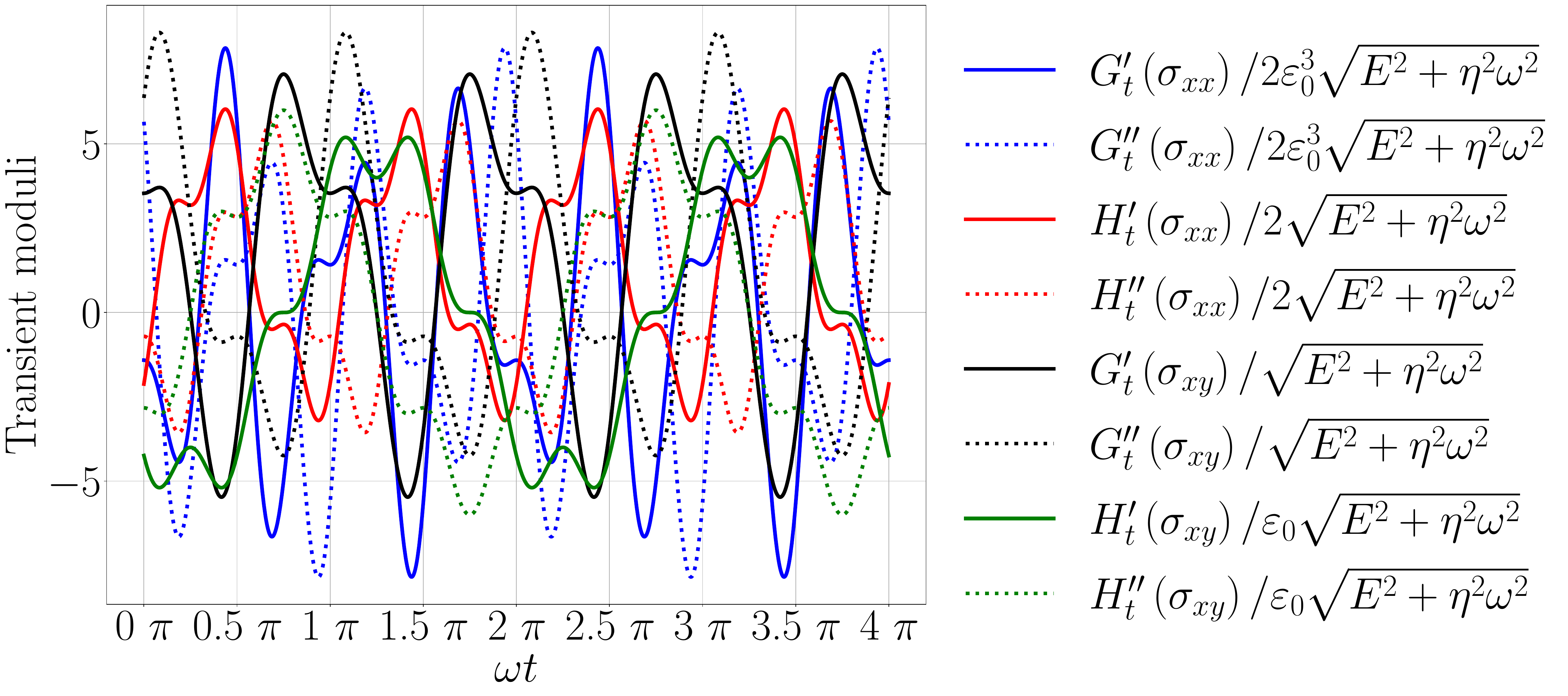}
	\caption{Evolution of the various moduli $G^{\prime}_t$, $G^{\prime\prime}_t$, $H^{\prime}_t$ and $H^{\prime\prime}_t$ of $\sigma_{xy}$ and $\sigma_{xx}$ divided, respectively, by $2\varepsilon_0^3\sqrt{E^2+\eta^2\omega^2}$, $2\sqrt{E^2+\eta^2\omega^2}$, $\sqrt{E^2+\eta^2\omega^2}$ and $\varepsilon_0\sqrt{E^2+\eta^2\omega^2}$ as a function $\omega t$ imposing $E/\eta\omega=\varepsilon_0=1$.}
	\label{fig:TimeModuliPiolaKelvin}
\end{figure*}

\begin{figure}
	\centering
	\includegraphics[width=\columnwidth]{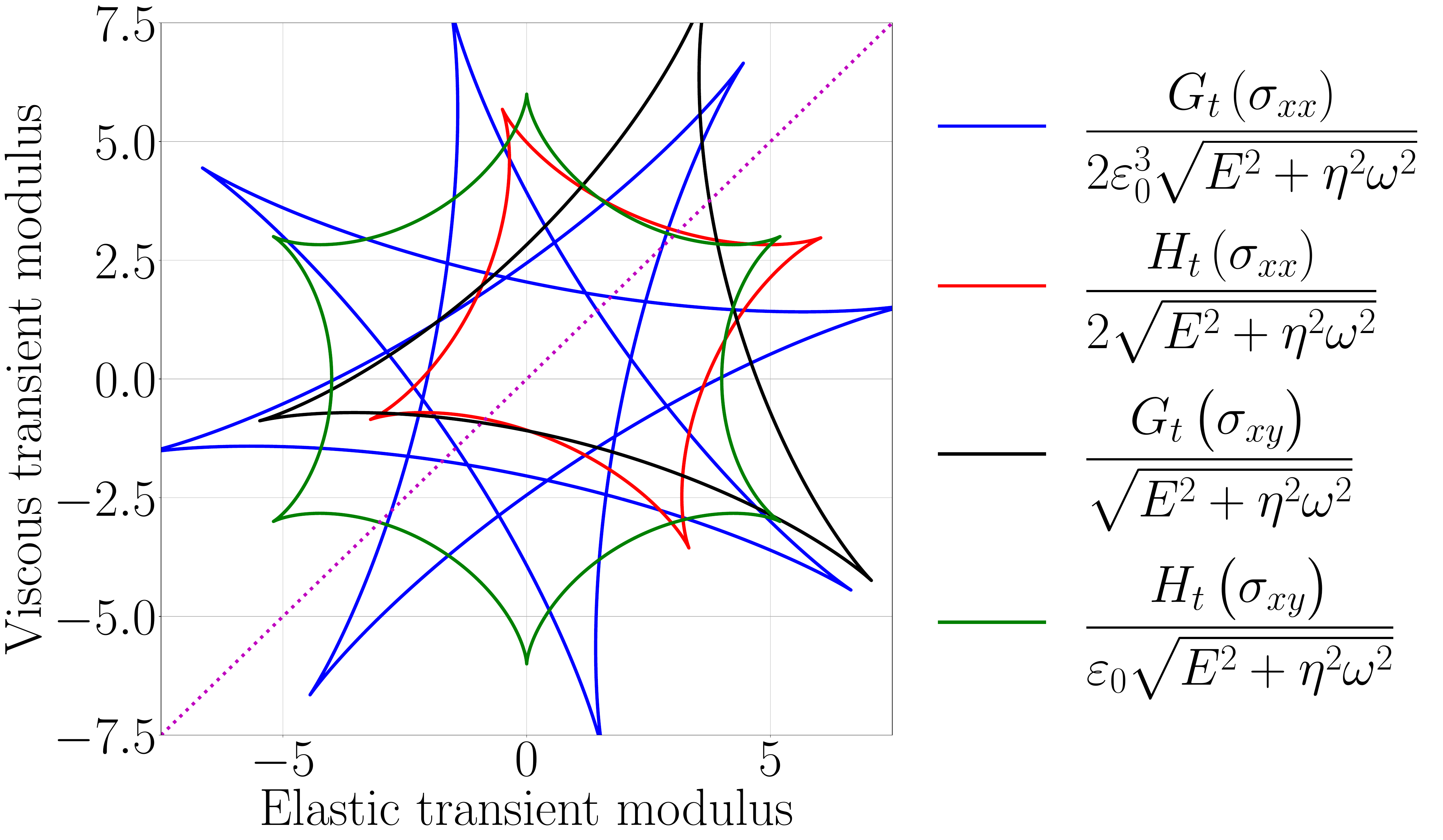}
	\caption{Evolution of the viscous moduli $G^{\prime\prime}_t$ and $H^{\prime\prime}_t$ of $\sigma_{xy}$ and $\sigma_{xx}$ divided, respectively, by $2\varepsilon_0^3\sqrt{E^2+\eta^2\omega^2}$, $2\sqrt{E^2+\eta^2\omega^2}$, $\sqrt{E^2+\eta^2\omega^2}$ and $\varepsilon_0\sqrt{E^2+\eta^2\omega^2}$ as a function of the elastic moduli $G^{\prime}_t$ and  $H^{\prime}_t$ of $\sigma_{xy}$ and $\sigma_{xx}$ divided, respectively, by $2\varepsilon_0^3\sqrt{E^2+\eta^2\omega^2}$, $2\sqrt{E^2+\eta^2\omega^2}$, $\sqrt{E^2+\eta^2\omega^2}$ and $\varepsilon_0\sqrt{E^2+\eta^2\omega^2}$ imposing $E/\eta\omega=\varepsilon_0=1$. The purple dotted line corresponds to the first bisector $y=x$.}
	\label{fig:PhaseModuliPiolaKelvin}
\end{figure}

Now if we look at the expressions for $\sigma_{xx}$ and $\sigma_{xy}$, we see that there are non zero contribution of each component. Normally, one would expect only the $\sigma_{xy}$ component to play a role but here, due to the increase of the strain amplitude $\varepsilon_0$, additional contributions appear. It is possible to recover the usual linear Kelvin-Voigt model expressed in terms of the Cauchy stress tensor $\boldsymbol{\sigma}$ when $\varepsilon_0\to0$. However, at high strain amplitude, there are variations in time of the various transient moduli. So here is the actual novelty of this approach because, with the steady-state approach, one would only find the average values over time. Now, one can find some more intricate evolutions to properly understand the behaviour of the materials. To give some illustrations to those equations, \cref{fig:TimeModuliPiolaKelvin} shows the time evolution of the various transient moduli of $\sigma_{xx}$ and $\sigma_{xy}$ with the parameters $\varepsilon_0=E/\eta\omega=1$. To give another perspective, the \cref{fig:PhaseModuliPiolaKelvin} gives the Cole-Cole plot, i.e. the viscous transient moduli as a function of the elastic transient moduli, with the same set of parameters. An interesting thing when looking at \cref{fig:PhaseModuliPiolaKelvin} is that, focusing on the black line for $G_t\left(\sigma_{xy}\right)$, which is usually observed in experiments presenting 3 peaks, there exist some portion of the cycle where either or both $G_t^{\prime}\left(\sigma_{xy}\right)$ and $G_t^{\prime\prime}\left(\sigma_{xy}\right)$ are negative. This is interpreted sometimes as sign of elastic recoil or negative dissipation. However, this example demonstrates that even for a model where the rheological parameters are properly defined without any doubts on thermodynamics, there is some strange behaviour happening which may lead skewed interpretations. In this case, a simple geometrical transformation of the Cauchy stress tensor $\boldsymbol{\sigma}$ into the second Piola-Kirchhoff stress tensor $\boldsymbol{S}$ leads to a rheological analysis which is much simpler due to the fact that the newly obtained moduli will be constant and corresponding exactly to a linear Kelvin-Voigt fluid. All these remarks are just a warning for people who are looking for a characterisation and an identification of their materials: some processing may be needed on experimental data to analyse properly the rheological behaviour having taken into account geometrical non linearities. 

We can carry out the overall analysis as in \cite{Rogers2017,Donley2019,Rogers2011,Rogers2012a,Rogers2012b} but we leave the rest of the comparison to the reader.

In general, what is interesting is that when $\varepsilon_0\gg1$, the main stress components are $\sigma_{xx}$ first due to the proportionality to $\varepsilon_0^4$, then $S_{yy}$, $\sigma_{xy}$ and $\sigma_{yy}$ with the proportionality to $\varepsilon_0^2$ and finally, the shear components $S_{xy}$ which is linear in $\varepsilon_0$.

\section{How to use this framework experimentally}\label{sec:Experiments}

All the previous demonstrations may seem really theoretical when compared to actual complex behaviour observed in experiments. Therefore, any experimental researcher may wonder how the understanding of rheology is enhanced by such framework. To answer this item, it is interesting to note that experiments are carried in \emph{real} time: that is to say, all the quantities are measured in the current configuration, omitting the conditions of the initial configuration. Hence, to rationalise the observed behaviour, it is relevant to come back to the initial configuration to follow what has happened since the beginning of the experiment both macroscopically and microspically. From this objective, concretely, for example, the stresses may be measured through any device to compute $\boldsymbol{\sigma}$ and come back to the \emph{original} stresses in the initial configuration $\boldsymbol{S}$ using the transformation given by \cref{eq:StressTensorPiola} in the case of a plan shear in the $xy$ plane. Hence, displaying the values of $\boldsymbol{S}$ instead of $\boldsymbol{\sigma}$ will bring a different perspective to the interpretation of stresses. Then, trying to find a relationship between $\boldsymbol{S}$ and $\boldsymbol{e}$ and its derivatives instead of $\boldsymbol{\varepsilon}$ and its derivatives can expand the field of possibilities and a broader understanding of the evolution of materials. It may not be as simple and confortable than a 1D variable like $\boldsymbol{\sigma}$ and $\boldsymbol{\varepsilon}$ but materials are dived in 3D space and inherit most of their properties from this initial description. Consequently, omitting this 3D dependence may lead to misunderstanding of the actual physics of rheology which was highlighted in this paper.

Furthermore, all the previous discussion can only happen when the experimental data are of high quality. Corrections of non linearities are indeed useful when there is a complete control of the materials behaviour in terms of data accuracy and data exhaustiveness
 
\section{Conclusion}\label{sec:Conclusion}

To put it in a nutshell, this paper has extended the usual Sequence of Physical Processes framework developed by \cite{Rogers2017} taking into account geometrical non linearities and geometrical corrections through the use of the second Piola-Kirchhoff stress tensor and the Green-Lagrange strain tensor. Applying the global framework to two very classical linear viscoelastic models showed how a simple rheological relation may bring much more complex observations in the current configuration. The last section highlighted this point inviting researchers investigating large amplitude oscillations to study the rheology of certain materials to properly account for geometrical non linearities before trying to conclude on certain observations.  

The main point of this paper is to point out that the initial choices of a representation, which can be the pure rheological model but also the intuition of the actual flow in a rheometer for example, are decisive. In this sense, starting with what can be seen \emph{a posteriori} as the wrong representation may force to carry on heavy artifacts which can be counterproductive. It must be remembered that the simplest descriptions, in terms of calculations and number of parameters, are usually the best to be able to interpret as simply as possible the very intricate phenomena we observe. 

A more general comment is that when we study large deformations in rheology or mechanics, heterogeneities in the deformations or in the flow profile may appear and lead to wrong interpretations. A closer look must happen to carefully pay attention to both the rheological measurements and the actual physics of the movement. 

\bibliography{Bibliographie_these}

\end{document}